\documentclass[sigconf]{acmart}

 \setcopyright{none}
 \renewcommand\footnotetextcopyrightpermission[1]{} 
 \pdfoutput=1

\usepackage{graphicx}
\usepackage{subcaption}

\usepackage{tabularx}

\usepackage{tikz}
\usepackage{multirow}
\usepackage{amsmath}
\usepackage{qcircuit}
\usepackage{algorithm}
\usepackage{algpseudocode}
\usepackage{graphicx}
\usepackage{caption}
\usepackage{standalone}
\usepackage{subcaption}
\usepackage{float}
\usepackage{braket}

\newcommand{\zx}[1]{{\color{magenta}{[ZX: #1]}}}

\usetikzlibrary{shapes,arrows,positioning}
\usetikzlibrary{shadows,arrows.meta,positioning,backgrounds,fit}
\tikzset{%
  materia/.style={draw, fill=blue!20, text width=6.0em, text centered, minimum height=1em,drop shadow},
  etape/.style={materia, text width=18em, minimum width=21em, minimum height=2em, rounded corners, drop shadow},
  etape2/.style={materia, fill = white, text width=14em, minimum width=14em, minimum height=2em, rounded corners},
  texto/.style={above, text width=6em, text centered},
  linepart/.style={draw, thick, color=black!50, -LaTeX, dashed},
  line/.style={draw, thick, color=black!50, -LaTeX},
  ur/.style={ text centered, minimum height=0.01em},
  back group/.style={fill=yellow!20,rounded corners, draw=black!50, dashed, inner xsep=10pt, inner ysep=10pt},
  back group2/.style={fill=white,rounded corners, draw=black!100, dashed, inner xsep=10pt, inner ysep=10pt},
}

\usetikzlibrary{external}
\begin{document}

\title{A comparison on constrain encoding methods for quantum approximate optimization algorithm}

\author{Yiwen Liu*}
\affiliation{%
  \institution{FinQ Tech Inc.}
  \country{USA}
}

\author{Qingyue Jiao*}
\affiliation{%
  \institution{University of Notre Dame}
  \country{USA}
}

\author{Yidong Zhou}
\affiliation{%
  \institution{Rensselaer Polytechnic Institute}
   \country{USA}
}

\author{Zhiding Liang}
\affiliation{%
  \institution{Rensselaer Polytechnic Institute}
   \country{USA}
}

\author{Yiyu Shi}
\affiliation{%
  \institution{University of Notre Dame}
  \country{USA}
}

\author{Ke Wan}
\affiliation{%
  \institution{FinQ Tech Inc.}
  \country{USA}
}

\author{Shangjie Guo}
\affiliation{%
  \institution{FinQ Tech Inc.}
  \country{USA}
}

\thanks{*These authors contributed to the work equally.}

\begin{abstract}

The Quantum Approximate Optimization Algorithm (QAOA) represents a significant opportunity for practical quantum computing applications, particularly in the era before error correction is fully realized. This algorithm is especially relevant for addressing constraint satisfaction problems (CSPs), which are critical in various fields such as supply chain management, energy distribution, and financial modeling. In our study, we conduct a numerical comparison of three different strategies for incorporating linear constraints into QAOA: transforming them into an unconstrained format, introducing penalty dephasing, and utilizing the quantum Zeno effect. We assess the efficiency and effectiveness of these methods using the knapsack problem as a case study. Our findings provide insights into the potential applicability of different encoding methods for various use cases.
\end{abstract}

\maketitle
\pagestyle{plain}

\section{Introduction}
\label{introduction}

Constraint Satisfaction Problems (CSPs) involve variables that must meet specific constraints. CSPs are crucial in artificial intelligence and operations research due to their ability to standardize various problems, such as the Maximum Cut problem, Knapsack problem, Eight Queens puzzle, Sudoku, and other logic puzzles. They find applications in fields like robotics, production planning, supply chain optimization, employee scheduling, project management, floor plan design, circuit layout, and network configuration.


Solving large CSPs can be computationally intense, especially for NP-complete problems. Researchers are now exploring quantum algorithms as a potential solution. Quantum computing can represent objective functions and potential solutions in superposition~\cite{l2024quantum, tan2023quct,wang2022quest,wang2024atomique}, allowing simultaneous exploration of all possibilities. This contrasts with classical methods that explore one or a few solutions at a time. Quantum optimization algorithms could greatly enhance the efficiency of solving CSPs.

In quantum optimization algorithms, the most notable algorithms and techniques include the QAOA~\cite{wan2023hybrid,liang2023hybrid,liang2024graph, zhuang2024improving}, quantum Annealing~\cite{morita2008mathematical}, and Adiabatic Quantum Computation (AQC)~\cite{albash2018adiabatic}. Also, Grover's algorithm can also be useful to tackle search-related tasks in certain optimization problems.

To effectively leverage quantum optimization algorithms, we need to properly encode constraint information and objective functions. In this paper, we review several alternative methods for encoding optimization problems and designing quantum circuits to solve constrained optimization problems, including quadratization, penalty dephasing, and the quantum Zeno effect.

\textbf{Review of Quadratization} 
Quadratization is a critical technique in quantum computing, particularly in the context of optimizing high-degree problems by converting them into quadratic forms, which are more amenable to quantum algorithms. This process is essential for various quantum mechanics applications, including quantum annealing and universal adiabatic quantum computing. The work by Dattani in 2019~\cite{dattani2019quadratization} provides a comprehensive exploration of Hamiltonian transformations that facilitate this conversion, demonstrating the utility of quadratization in discrete optimization problems, crucial for efficient quantum computations.

Further advancements in quadratization methods have been made to improve the accuracy and efficiency of quantum algorithms. Bychkov introduced novel algorithms and software capabilities for quadratization, focusing on non-autonomous dynamical systems of arbitrary dimensions~\cite{bychkov2024exact}. Their work provides a robust framework for applying quadratization across various disciplines, showcasing its versatility and effectiveness in handling complex optimization tasks.

Another significant contribution to this field is the introduction of the Relaxed-Energy Quadratization method~\cite{zhao2021revisit}, which addresses the inconsistencies of the traditional energy quadratization approach. This method enhances both accuracy and consistency while maintaining the beneficial properties of the original approach, making it a powerful tool for solving gradient flow models and other nonlinear optimization problems in quantum computing.

These research efforts collectively advance the understanding and application of quadratization in quantum computing, enabling more efficient and accurate solutions to complex optimization problems that are central to the field.

\textbf{Review of penalty dephasing} 
The penalty dephasing is a well-studied approach in quantum circuit design, known for its effectiveness in addressing non-Ising target functions. Rooted in AQC and the adiabatic approximation, this algorithm involves gradually evolving the Hamiltonian from an initial state to a final state that encodes the optimization problem \cite{aharonov2008adiabatic}\cite{albash2018adiabatic}. The system, initially prepared in the ground state of the initial Hamiltonian, ideally remains in this state throughout the evolution, reaching the optimal solution. AQC has proven effective for various optimization problems, including Ising spin glasses and satisfiability problems.

Any optimization problem can be reformulated for AQC by constructing a Hamiltonian whose ground state encodes the solution. The quantum system is initially prepared in the ground state of a simple Hamiltonian and evolves adiabatically under a time-dependent Hamiltonian to reach the final Hamiltonian, which encodes the optimal solution.

In related research, Farhi et al. (2014) \cite{farhi2014quantum} proposed a quantum algorithm for generating approximate solutions to combinatorial optimization problems. The accuracy of the algorithm improves as the integer $p$ increases. The quantum circuit that implements the algorithm consists of unitary gates whose locality is no more than that of the objective function for which the optimal value is sought. Grand'rive et al. (2019) \cite{de2019knapsack} introduced a relaxed algorithm for solving constrained optimization problems with non-standard Ising target functions. This approach adds penalties for solutions that do not meet the constraints, implemented through separate "cost addition" and "constraint testing and penalty dephasing" blocks, rather than embedding them within the classical Ising Hamiltonian. Building on this work, Lukas et al. (2021) \cite{lucas2021ibmsolution} developed a robust implementation that further advances these techniques.

To enable these circuits, quantum networks capable of performing elementary arithmetic operations, including modular exponentiation, were introduced by Vedral et al. (1995) \cite{vedral1996quantum}. Additionally, various quantum adders have been developed, such as the quantum ripple-carry adder by Cuccaro et al. (2004) \cite{cuccaro2004new}, the quantum Fourier transform adder by Draper (2000) \cite{draper2000addition}, and the adder by Ruiz et al. (2014) \cite{ruiz2017quantum}. Modular exponentiation is a critical operation that dominates both the time and memory complexity in Shor's quantum factorization algorithm.

\textbf{Review of Quantum Zeno:} 
The quantum Zeno effect, also known as the Turing paradox, is a phenomenon in quantum mechanical systems where continuous measurement or observation prevents a quantum system from evolving, causing it to remain in its initial state. This effect can slow down the time evolution of a particle by frequently measuring it in a specific basis, thereby preventing transitions to states different from its initial state. The quantum Zeno effect arises from the fundamental properties of the Schrödinger equation and is unique to quantum systems, potentially providing novel solutions to a variety of problems.

The quantum Zeno effect has been extensively studied in the literature. Facchi et al. (2009) \cite{facchi2008quantum} outline three methods for achieving the Zeno effect: projective measurements, unitary kicks, and strong continuous coupling. They also introduce the concept of Quantum Zeno subspaces through blockwise decomposition. Burgarth et al. (2020) \cite{burgarth2020quantum} explore the evolution of a finite-dimensional quantum system under frequent kicks, where each kick is a generic quantum operation. They develop a generalization of the Baker-Campbell-Hausdorff formula, revealing an adiabatic evolution. Herman et al. (2022) \cite{herman2022portfolio} propose a technique that utilizes quantum Zeno dynamics to solve optimization problems with multiple arbitrary constraints, including inequalities. They demonstrate that the dynamics of quantum optimization can be efficiently confined to the in-constraint subspace through repeated projective measurements. To address the limitations of adiabatic quantum computation, Yu et al. (2022) \cite{yu2021quantum} adopt a Zeno method in quantum simulated annealing, where a series of eigenstate projections are employed. In their approach, the path-dependent Hamiltonian is augmented by a sum of Pauli X terms, whose contributions vanish at the beginning and end of the path.

In this paper, we propose a formal application of the quantum Zeno effect to solve optimization problems, significantly broadening its potential applications. The quantum Zeno effect allows the total Hilbert space to be partitioned into Zeno subspaces, where different components of the density matrix can evolve independently within each sector. By continuously measuring the system, the initial state within one of these subspaces maintains a survival probability of unity. We extend this concept to the QAOA by restricting system evolution to the subspace where feasible solutions reside, rather than embedding all constraints directly into the target function. 

The rest of the article is structured as follows: in the Materials and Methods section, we provide a detailed mathematical explanation and implementation of the three methods. In the Results and Analysis section, we evaluate the performance of these methods on knapsack problems with varying problem sizes and parameters.

\section{Methods}
\label{materialsandmethods}

\subsection{Quantum Approximate Optimization Algorithm}
QAOA\cite{farhi2014quantum} is a hybrid quantum-classical algorithm widely used to find approximate solutions to combinatorial optimization problems. specifically, QAOA involves the five main steps: problem setup, unitary transformation, ansatz construction, classical optimization, and measurement.

In the first problem setup, we start with a problem that we want to optimize and encode the optimization problem's objective function into a cost Hamiltonian \( H_C \). The goal is to find the ground state of the cost Hamiltonian \(H_C\). Then, unitary transformations are defined by
   \[ U(H_C, \gamma) = e^{-i \gamma H_C}, \quad U(H_B, \beta) = e^{-i \beta H_B} \]
    
where:
    \begin{itemize}
        \item \( H_C\) encodes the optimization problem (objective function).
        \item \( H_B \) is a mixing Hamiltonian. 
        \item Real parameters $\gamma$ and $\beta$ embedded in ansatz state need to classically optimized.
    \end{itemize}
    \
    \
In ansatz construction, we need to create an ansatz state ($| \psi(\vec{\gamma}, \vec{\beta}) \rangle $) by starting from an initial quantum state \( |\psi_0\rangle \) and then applying a series of unitary transformations  \( U(H_C, \gamma) \) and \( U(H_B, \beta) \) on this initial state as below:
     \[ |\psi(\vec{\gamma}, \vec{\beta})\rangle = U(H_B, \beta_p) U(H_C, \gamma_p) \ldots U(H_B, \beta_1) U(H_C, \gamma_1) |\psi_0\rangle\]
where $p$ donotes total number of QAOA layers. Next, we use classical algorithms to optimize parameters $\vec{\gamma} = [\gamma_1, \gamma_2, ...\gamma_p]$ and $\vec{\beta} = [\beta_1, \beta_2, ...\beta_p]$ to minimize the expectation value of the optimization problem Hamiltonian \(\langle H_C \rangle =  \langle \psi(\vec{\gamma}, \vec{\beta}) | H_C | \psi(\vec{\gamma}, \vec{\beta}) \rangle \). In this work, we used Constrained Optimization BY Linear Approximation (COBYLA) algorithm designed in scipy library. 

Finally, the quantum state \(| \psi(\vec{\gamma}, \vec{\beta}) \rangle \) is measured and outputs a classical bit string. This bit string represents an approximate solution to the original optimization problem.

In the following sections, we will introduce three different approaches for encoding optimization problems with constraints.

\subsection{Constraint Encoding Methods for QAOA}

\begin{figure}[t!]
    \begin{subfigure}[c]{0.23\textwidth}
        \raggedleft

        \centering
        \resizebox{\textwidth}{!}{
        \Qcircuit {
            \lstick{\ket{0}^{\otimes m}} & \gate{H}\ar@{.}[]+<2em,1em>;[d]+<2em,-1em>
 & \multigate{1}{e^{-i\gamma H_C}} & \multigate{1}{e^{-i\beta H_B}}\ar@{.}[]+<3em,1em>;[d]+<3em,-1em> & \qw \\
            \lstick{\ket{0}^{\otimes c}} & \gate{H} & \ghost{e^{-i\gamma H'_C}} & \ghost{e^{-i\beta H'_B}} & \qw \\
            }
        }
        \caption{\small Quadratization (QUBO)}
        \label{fig:QAOAstructure}
    \end{subfigure}
    \begin{subfigure}[c]{0.45\textwidth}  
        \centering
        \resizebox{\textwidth}{!}{
        \Qcircuit {
            \lstick{\ket{0}^{\otimes m}} & \gate{H}\ar@{.}[]+<2em,1em>;[d]+<2em,-6em> & \gate{U(C,\gamma)} & \multigate{1}{\text{adder}} & \qw & \gate{\begin{array}{c}\text{penalty} \\ \text{dephasing}\end{array}} & \multigate{2}{\begin{array}{c}\text{reinitia-} \\ \text{lization}\end{array}}  & \gate{U(B,\beta)}\ar@{.}[]+<3em,1em>;[d]+<3em,-6em> & \qw \\
            \lstick{\ket{0}^{\otimes n}} & \qw & \qw & \ghost{\text{adder}} & \multigate{1}{\begin{array}{c}\text{constraint} \\ \text{testing}\end{array}} & \qw & \ghost{\begin{array}{c}\text{reinitia-} \\ \text{lization}\end{array}} & \qw & \qw \\
            \lstick{\ket{0}} & \qw & \qw & \qw & \ghost{\begin{array}{c}\text{constraint} \\ \text{testing}\end{array}} & \ctrl{-2}  & \ghost{\begin{array}{c}\text{reinitia-} \\ \text{lization}\end{array}} & \qw & \qw \\
            }
        }
        \caption[]{\small Penalty dephasing}
        \label{fig:dephasingstructure}
    \end{subfigure}
    \begin{subfigure}[c]{0.42\textwidth}   
        \centering
        \resizebox{\textwidth}{!}{
        \Qcircuit {
            \lstick{\ket{0}^{\otimes m}} & \gate{H}\ar@{.}[]+<2em,1em>;[d]+<2em,-5em> & \gate{U(C,\gamma)} & \multigate{1}{\text{adder}} & \qw  & \multigate{1}{\begin{array}{c}\text{reinitia-} \\ \text{lization}\end{array}}  & \gate{U(B,\beta)}\ar@{.}[]+<3em,1em>;[d]+<3em,-5em> & \qw \\
            \lstick{\ket{0}^{\otimes n}} & \qw & \qw & \ghost{\text{adder}} & \multigate{1}{\begin{array}{c}\text{constraint} \\ \text{testing}\end{array}}  & \ghost{\begin{array}{c}\text{reinitia-} \\ \text{lization}\end{array}} & \qw & \qw \\
            \lstick{\ket{0}} & \qw & \qw & \qw & \ghost{\begin{array}{c}\text{constraint} \\ \text{testing}\end{array}} & \meter & \lstick{\ket{0}} &\qw  \\
            }
        }  
        \caption[]{\small Quantum Zeno}
        \label{fig:zenostructure}
    \end{subfigure}
    \caption[]{\small One layer circuit structures for constraint encoded variational quantum algorithms: (a) Quadratization QAOA, (b) Penalty dephasing QAOA, and (c) quantum Zeno QAOA. We used Hadamard array for initialization. Vertical dashed lines indicates one QAOA layer.} 
    \label{fig:qcircuits}
\end{figure}
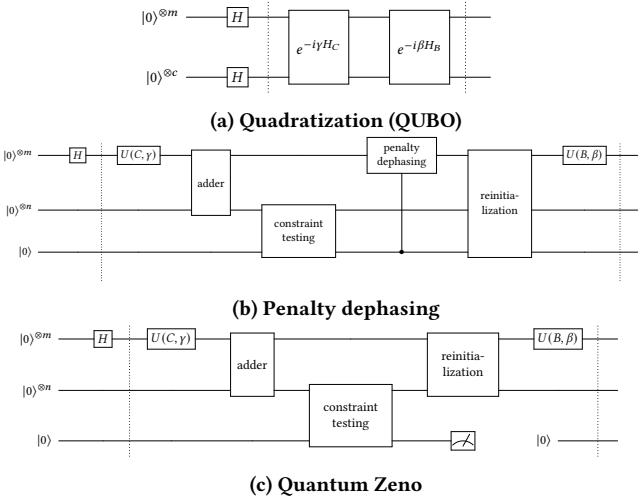

In this section we discuss how quantum circuits are designed for each encoding methods, as shown in figure~\ref{fig:qcircuits}. There are seventeen distinct special classes of linear constraints \cite{miplib2017}. We choose the most of the special classes which can be reformed into a knapsack type of constrains. The knapsack problem can be summarized as following. Let there be total of $m$ items with index $i = 1,\ 2,\ ...,\ m$ choices for each item $x_i \in \{0, 1\}$, we aim to maximize total value
\begin{align}
V = \sum_{i=1}^m v_ix_i, \quad \text{subject to }\quad \sum_{i=1}^m w_ix_i\leq W,
\end{align}
where $w_i \in \mathbb{N}$ is the weight of individual item, $W \in \mathbb{N} \geq 2 $ is the maximum weight limit, $v_i \in \mathbb{N}$ is the value of individual item.

For all three encoding methods, the first $m$ qubits are dedicated to encoding the decision binary variable $x_{i}$ $\forall i \in [1, m]$.  $x_{i} = 1$ indicates the selection of ith knapsack, and $x_{i} = 0$ otherwise. Consequently, each vector $x = [x_{1}, ..., x_{m}]$ represents a distinct selection scenario.

To calculate the number of ancillary qubits required to represents the constrains term, we denote $c = \left\lceil \log_2(W) \right\rceil + 1$, where $\lceil\rceil$ denote for ceil function. In the case that $\exists k \in \mathbb{N},\ W = 2^k$, we can use the last ancilla qubit to indicate if the constrain is satisfied. And in the case where $W$ is not powers of 2 but an arbitary integer, we can use a trick to add a constant such that it can be treated as a powers of 2:
\begin{align}
\sum_{i=1}^m w_ix_i  < W  \Leftrightarrow \sum_{i=1}^m w_ix_i  + s < W + s = 2^{c} 
\end{align}

We can turn value \(2^c\) into binary representation\(\sum_{k} 2^k c_k\), where \(k\) is the number of binary digits needed. Upon introducing a slack variable, the problem is equivalent to minimizing the following cost function:
\begin{align}
f = -\sum_{i=1}^m v_ix_i + P(\sum_{i=1}^m w_ix_i + \sum_{k=0}^c 2^kc_k - W)^2
\label{eqn:QUBO}
\end{align}
where $P\in \mathbb{R}$ is the Lagrange multiplier determining the importance of our constrains.

\begin{figure*}[h!]
    \centering
    \begin{subfigure}[t!]{0.22\textwidth}
        \centering
        \includegraphics[width=\textwidth]{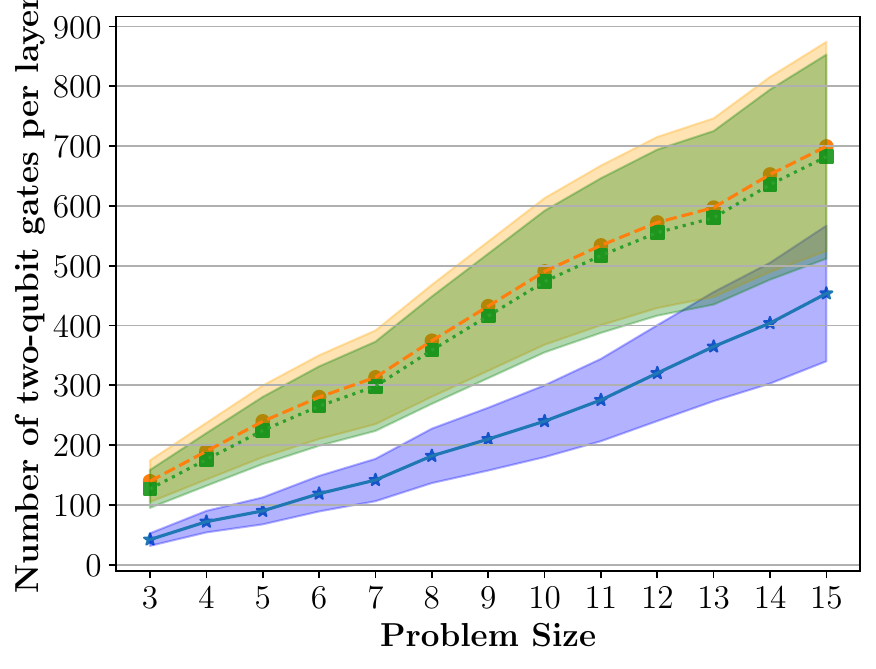}    
        \label{fig:size_change_quantum_resource_2qgate}
        \caption[]{\small No. two-qubit gates per layer} 
    \end{subfigure}
    \hspace{0.015\textwidth}
    \begin{subfigure}[t!]{0.22\textwidth}  
        \centering 
        \includegraphics[width=\textwidth]{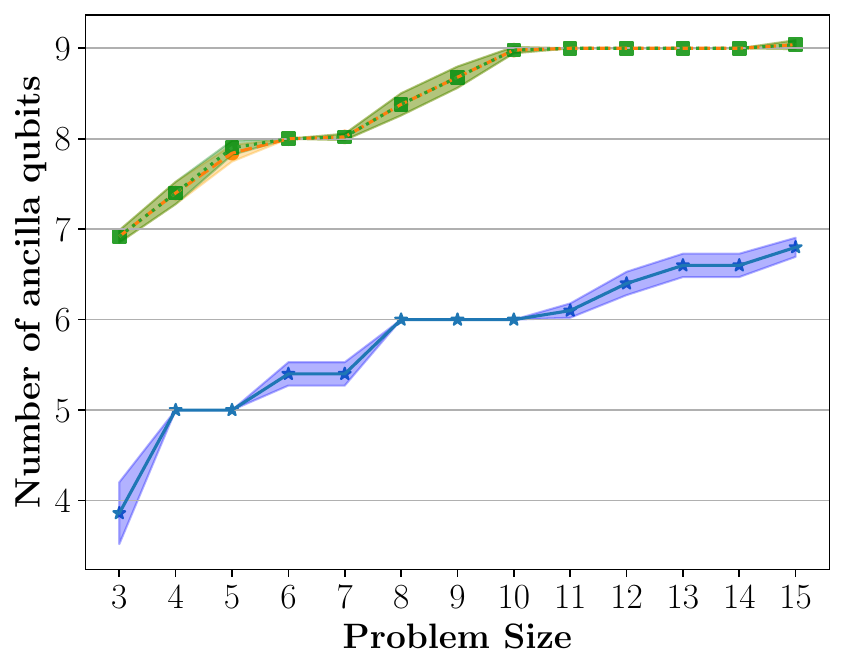}  
        \label{fig:size_change_quantum_resource_ancilla}
        \caption[]{\small No. ancilla qubits}
    \end{subfigure}
    \hspace{0.015\textwidth}
    \begin{subfigure}[t!]{0.22\textwidth}   
        \centering 
        \includegraphics[width=\textwidth]{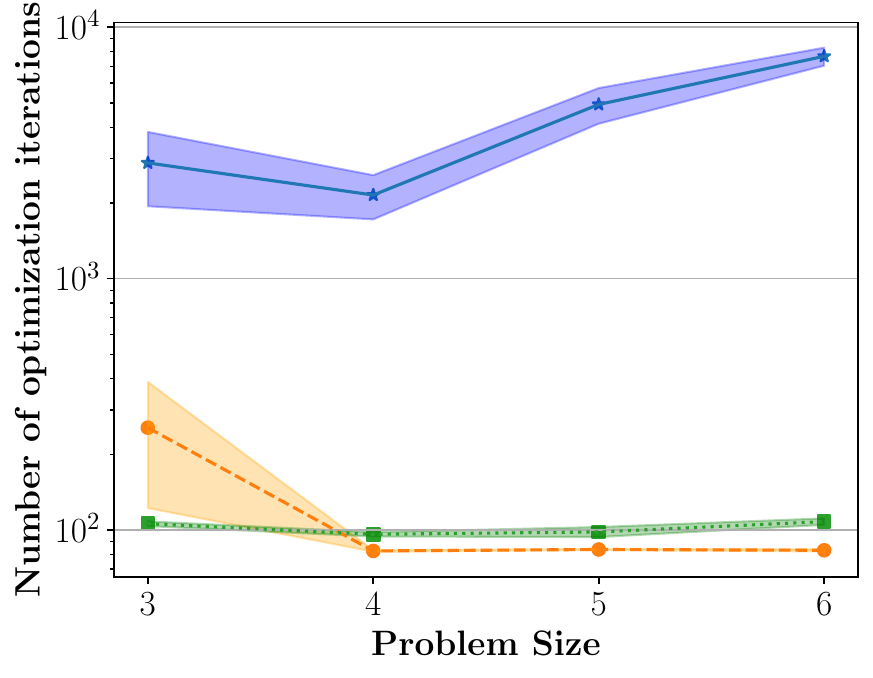}    
        \label{fig:size_change_quantum_resource_nfev}
        \caption[]{\small No. optimization iterations}
    \end{subfigure}
    \hspace{0.015\textwidth}
    \begin{subfigure}[t!]{0.22\textwidth}   
        \centering 
        \includegraphics[width=\textwidth]{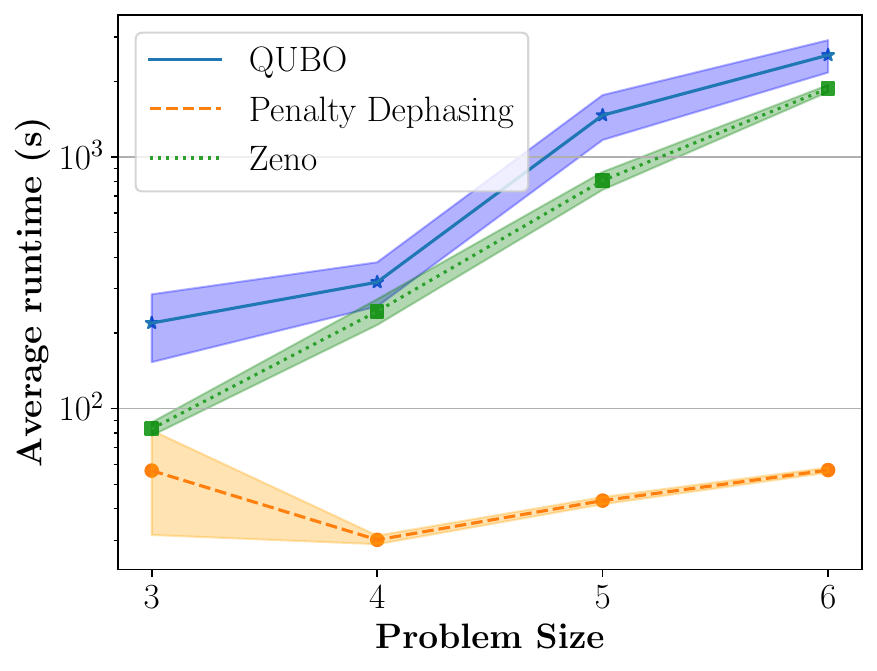}   
        \label{fig:size_change_quantum_resource_runtime}
        \caption[]{\small Average simulation runtime}
    \end{subfigure}
    \caption[]{\small Quantum resources required for different problem sizes. (a) Number of CNOT gate for each QAOA layer after decomposition; (b) Number of ancilla qubits needed in addition to data quantum register; (c) Number of optimization iterations for classical optimizer to converge; (d) Total run time for classical simulation of the QAOA algorithm. All plot are generated averaging over 50 random problems with a fixed number of QAOA layers $p$ = 5. The shaded area denotes the distribution of 25\%-75\% of samples. Y-axes in figures (c) and (d) are drawn on a log scale for better illustration.} 
    \label{fig:quantum_metrics_varying_size}
\end{figure*}

    
    
    

\subsubsection{Quadratization}
The first approach we tested is to translate CSP into QUBO problem by relaxing the linear constrains to quadratic terms into objective functions. For the knapsack problem, we form a new unconstrained objective function as below: 
Next we can map QUBO  objective function~\ref{eqn:QUBO} into an Ising Hamiltonian with form:
\begin{eqnarray}
H'_C&=& \sum_{a}Q_{ij}\frac{(\mathbb{I}+Z)_i}{2}\frac{(\mathbb{I}+Z)}{2} + \sum_{i=1}^m B_i\frac{(\mathbb{I}+Z)_i}{2}
\end{eqnarray}
by performing the mapping: $
x_i\rightarrow \frac{1}{2}(\mathbb{I}+Z)_i,\quad i=1,...,n
$, where $Z$ stands for Pauli Z operator. And matrix $Q$ and vector $B$ is calculated by:
\begin{equation}
Q_{ij} =
\begin{cases} 
2 w_i w_j P & \text{if } 1 \leq i < j \leq m \\
2^{i + j - 2m + 1} P & \text{if } m+1 \leq i < j \leq m + c \\
w_i 2^{j-m+1} P & \text{if } 1 \leq i \leq m \text{ and } m+1 \leq j \leq m + c
\end{cases}
\end{equation}
and 
\begin{equation}
B_i = 
\begin{cases}
-v_i + P (w_i^2 - 2 W w_i) & \text{if } 1 \leq i \leq m \\
P [4^{(i-m)} - W 2^{(i-m+1)}] & \text{if } m+1 \leq i \leq m + c
\end{cases}
\end{equation}

To simplify this we have used of the fact that $Q_{ij}$ is diagonal symmetric in the first term, and $Z_i$ commutes with $Z_j$.
And we dropped the last term is a constant term, with no effect on the eigenstates. We are only interested in constructing a quantum circuit corresponding to matrices Q and B for implementing $\exp(-i\gamma H'_C)$.

To construct $H'_B$ for implementing the operator $\exp(-i\beta H'_B)$, we use the Pauli X operator array $H'_B = \sigma_x^{\otimes m+c}$, where $m+c$ is the number of qubits needed for encoding the full objective function. The overall quantum circuit for quadratization QAOA is shown in figure~\ref{fig:QAOAstructure}.



\subsubsection{Penalty dephasing}

The second approach to encoding the constraint in the Knapsack problem utilizes penalty dephasing\cite{de2019knapsack}. In penalty dephasing, the full objective function is divided into two distinct components: the return part and the constraint penalty part. The return part consists of the original objective function:
$f_\text{return} = -\sum_{i=1}^m x_iv_i$
and a penalty phase will be added to a solution $\vec{x}$ when
$\sum_{i=1}^m w_ix_i - W > 0$. The penalty dephasing circuit is comprised of six main components, as shown in Figure~\ref{fig:dephasingstructure}.

The first block encodes the return part. The value of each individual knapsack $v_i$ is normalized and reflected by the phase of each corresponding qubit. With phase gate
\begin{equation}
\text{P}_i(\lambda) = \left[ \begin{array}{cc} 1 & 0 \\ 0 & e^{i\lambda} \end{array} \right]
\end{equation}
for the $i$-th qubits, then the return operator $U(H_C, \gamma)$ is can be represented as 
\begin{equation}
\otimes_{i=1}^{n} \text{P}_i\left(-\gamma * v_i\right).
\end{equation} 

The second block constructs a weight adder to verify the knapsack constraint. Once a certain knapsack selection scenario is reached, the cost, the total weight of selected knapsacks in this case, is calculated by the weight adder. Ancillary qubits are allocated for the addition operation. The number of ancillary qubits $n = \left\lceil \log_2(\sum_{i=1}^m w_i) \right\rceil + 1$ should be sufficient to represent binary decomposition of summation of all item weights.

The adder unitary is defined such that:
\begin{equation}
\text{ADD}\left|x_{1}, ..., x_{n}\right\rangle\otimes \left|0\right\rangle := \left|x_{1}, ..., x_{n}\right\rangle \otimes \left|\sum_{i=1}^m w_i x_i\right\rangle,
\end{equation}
which can be simplified as
\begin{equation}
\text{ADD}\left|\vec{x}\right\rangle = \left|\vec{x}\right\rangle \otimes \left|\text{Weight}(\vec{x})\right\rangle.
\end{equation}
where $|\text{Weight}(\vec{x})\rangle $ is the binary decomposition of the total weight by m ancillary qubits. For implementation, We use quantum Fourier transform (QFT) adder to add weight for all senario.~\cite{ruiz2017quantum}

The third block of the circuit verifies the satisfaction of the constraints and records the results in a flag qubit, which is defined as:
\begin{equation}
\text{TEST}\left|\text{Weight}(\vec{x})\right\rangle \otimes \left|0\right\rangle = \left|\text{Weight}(\vec{x})\right\rangle \otimes \left|\text{if}\ \text{Weight}(\vec{x})>W  \right\rangle.
\end{equation}
Such that flag qubit indicates a constraint violation after the constrain testing protocol. 

The fourth block of the circuit imposes a penalty on all infeasible solutions by applying controlled phase gates to all qubits in the first block. The control qubit is the flag qubit. The penalty dephasing effectively creates the inference and decreases the likelihood of infeasible solutions. The magnitude of the penalty phase is determined by the extent to which the weight exceeds the limit:
\begin{equation}
\text{PENAL}\left|\psi\right\rangle \otimes \left|a\right\rangle = e^{-ia\alpha}\left|\psi\right\rangle \otimes \left|a\right\rangle
\end{equation}
where $a\in \{0,1\}$ is valur of flag qubit and $\alpha $ is a positive, tunable parameter representing the strength of penalty. 

The fifth block reinitializes all ancilla registers to prepare for the next calculation, i.e.
\begin{equation}
\text{REINIT} = (I\otimes\text{TEST}^\dagger)(\text{ADD}^\dagger\otimes I).
\end{equation}

The final mixer operator $U(H_B, \beta) = \exp(-iH_B\beta)$ is implemented with single-qubit X gates as the previous approach, but just on data register, which is expressed as $H_B = \sigma_x^{\otimes m} $.

\begin{figure*}[h!]
    \centering
    \begin{subfigure}[t!]{0.25\textwidth}
        \centering
        \includegraphics[width=\textwidth]{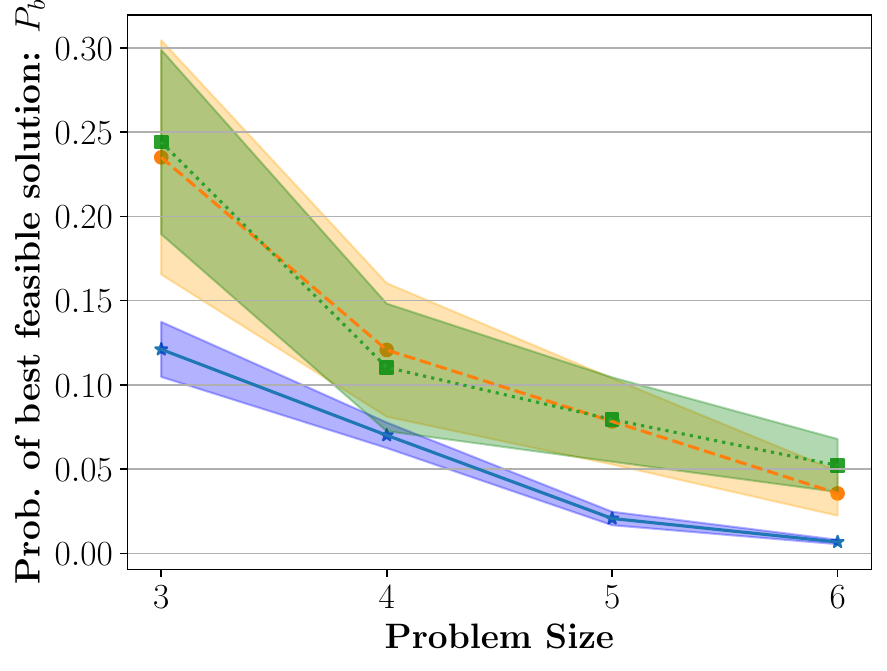}  
        \label{fig:p_change_performance_Pb}
        \caption[]{\small Prob. of best feasible solution} 
    \end{subfigure}
    \hspace{0.015\textwidth}
    \begin{subfigure}[t!]{0.25\textwidth}  
        \centering 
        \includegraphics[width=\textwidth]{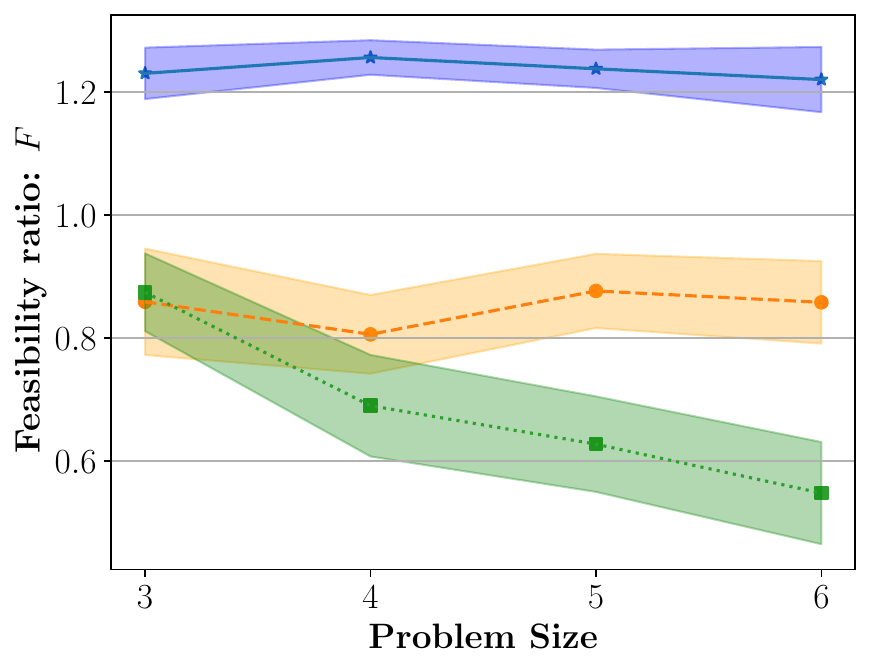} 
        \label{fig:p_change_performance_F}
        \caption[]{\small Feasibility ratio}
    \end{subfigure}
    \hspace{0.015\textwidth}
    \begin{subfigure}[t!]{0.25\textwidth}   
        \centering 
        \includegraphics[width=\textwidth]{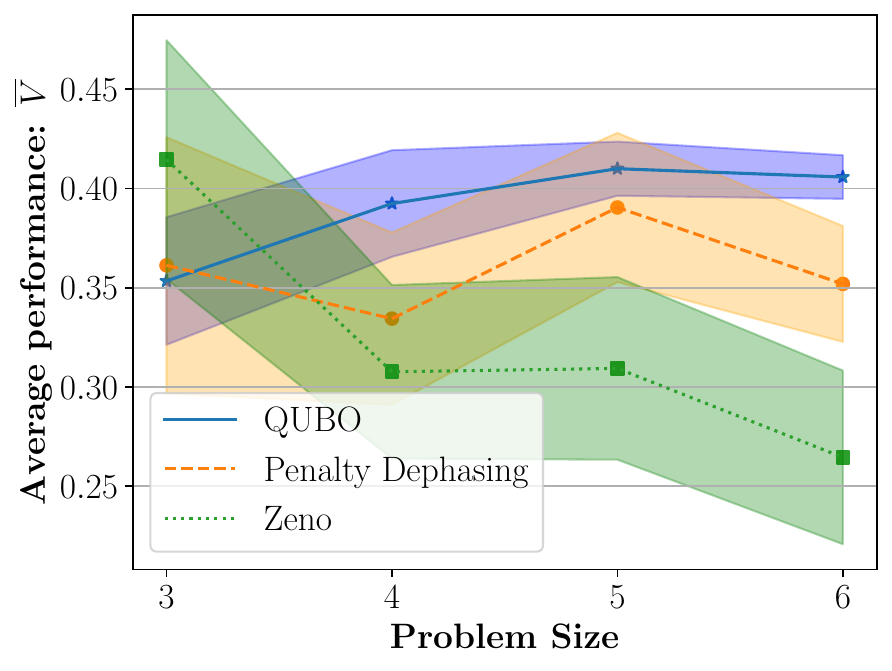}    
        \label{fig:p_change_performance_V}
        \caption[]{\small Average performance}
    \end{subfigure}
    \caption[]{\small Numerical performance for problem sizes from 3 to 6 with fixed number of layers $p=5$, measure by different metrics: (a) statistical probability of best feasible solution $P_B$; (b) Feasibility ratio $F$; and (c) Average performance $V$. The shaded area denotes the distribution of 25\%-75\% of samples.} 
    \label{fig:performance_varying_size}
\end{figure*}

\subsubsection{Quantum Zeno}
This approach is inspired by the fact that frequent measurements on a quantum system can project the system onto a reduced subspace. By frequently measuring the flag qubit, which indicates whether the solution (in superposition) is feasible or not, we can potentially confine the quantum evolution within a quantum Zeno subspace. This behavior, in turn, may effectively enforce the constraint, ensuring that it remains satisfied as the quantum state evolves \cite{Herman_2023}.

To illustrate how quantum Zeno effect works, let us examine a quantum system in a finite-dimensional Hilbert space $ \mathbb{H}$ governed by the unitary operator $U(t)=e^{-iHt}$, where $H$ is the time-independent Hamiltonian of the system. 
Assume a weak measurement is represented by a set of orthogonal projection operators $P$, which projects the time evolution on to the subspace where all constraints are satisfied. The measurement determines whether the system is in the subspace  $P\mathbb{H}$.  If the initial density matrix of the system is set as $\rho_{\boldsymbol{0}}=\lvert \psi_0 \rangle \langle\psi_0\rvert$, where $\lvert \psi_0 \rangle$  belongs to the subspace $ P\mathbb{H} $, the state of the system at time $t$ is given by:
$$\rho_{\boldsymbol{t}} = U(t) \rho_{\boldsymbol{0}} U^\dagger(t)$$

Assume that $P$ and $H$ does not commute, then the state no longer belongs to $ P\mathbb{H} $. If we measure with projector $P$ then
$$\rho_{\boldsymbol{t}}\rightarrow P\rho_{\boldsymbol{t}}P=P U(t) \rho_{\boldsymbol{0}} U^\dagger(t)P  =V(t)\rho_{\boldsymbol{0}} V^\dagger(t)$$
where $V(t) \equiv PU(t)P$. Next we examine the case where we measure $N$ times with $P$ during period of time $t$:
$$\rho^{(N)}_{\boldsymbol{t}} =V_{\boldsymbol{N}}(t)\rho_{\boldsymbol{0}}V^\dagger_{\boldsymbol{N}}(t) \quad\quad V_{\boldsymbol{N}}(t)\equiv[PU(t/N)P]^N$$
in the limit $N\to\infty$, we can Taylor expand $U(t)$ and drop the second and higher order terms:

\begin{align*}
    \lim_{N\to\infty}V_{\boldsymbol{N}}(t) &=\lim_{N\to\infty}[Pe^{-iH(t/N)}P]^N\\ 
    &=\lim_{N\to\infty}\left[ P \left(1 - iH\frac{t}{N} + O\left(\frac{1}{N^2}\right)\right) P\right]^N \\
    &= e^{-iPHPt}\\
\end{align*}

From the calculation, we conclude that the Zeno Hamiltonian projected on $P\mathbb{H}$ subspace: $H_{\boldsymbol{Z}}\equiv PHP$, where the time evolution is constrained to.

When constructing the quantum circuit for our knapsack optimization problem, we leverage the subroutines in penalty dephasing including return, adder, constraint testing, and mixer. For reinitilazation we only need to revert adder such that: $\text{REINIT} = \text{ADD}^\dagger$. And constrain enforcing with Zeno, we take one measurement at the flag qubit and reprepare it to $|0\rangle$. Note that we do not post-select based on these extra measurement results, since this post-selection dramatically increase the number of shots for quantum Zeno and we would like to compare each encoding methods fairly.


\begin{figure*}[h!]
    \centering
    \begin{subfigure}[t]{0.22\textwidth}
        \centering
        \includegraphics[width=\textwidth]{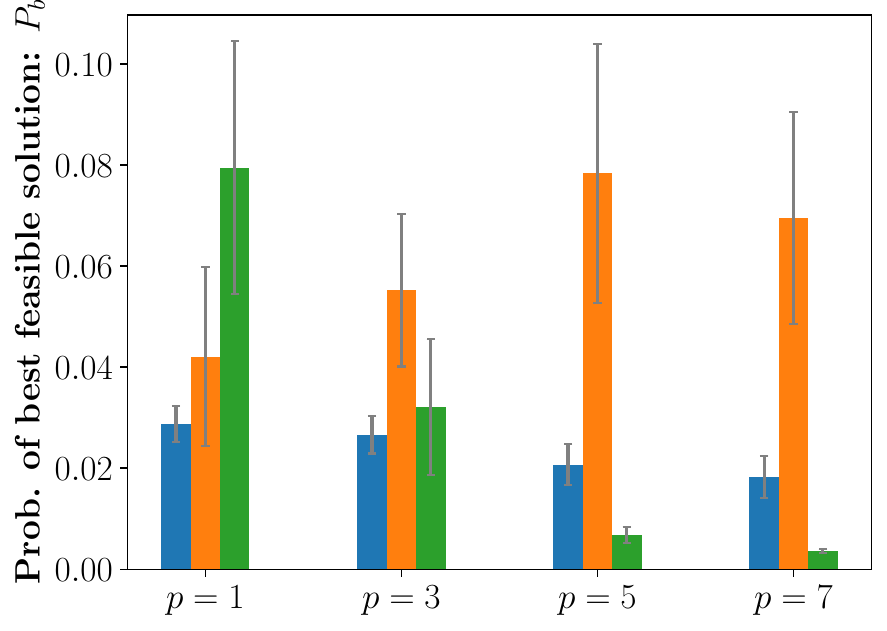}
        \label{fig:_p_change_performance_Pb}
        \caption[]{\small Prob. of best solution} 
    \end{subfigure}
    \hspace{0.02\textwidth}
    \begin{subfigure}[t]{0.22\textwidth}  
        \centering 
        \includegraphics[width=\textwidth]{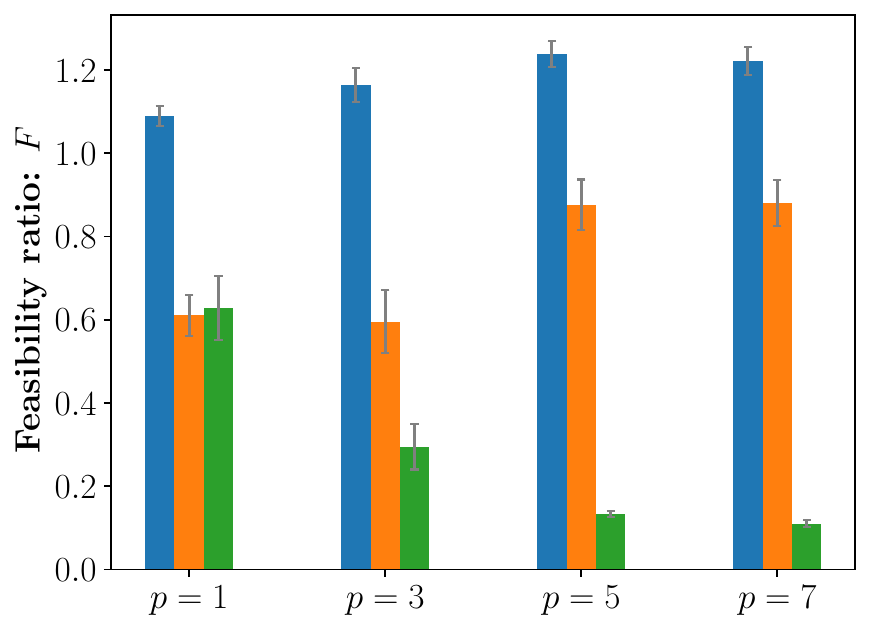}  
        \label{fig:p_change_performance_F}
        \caption[]{\small Feasibility ratio}
    \end{subfigure}
    \hspace{0.02\textwidth}
    \begin{subfigure}[t]{0.22\textwidth}   
        \centering 
        \includegraphics[width=\textwidth]{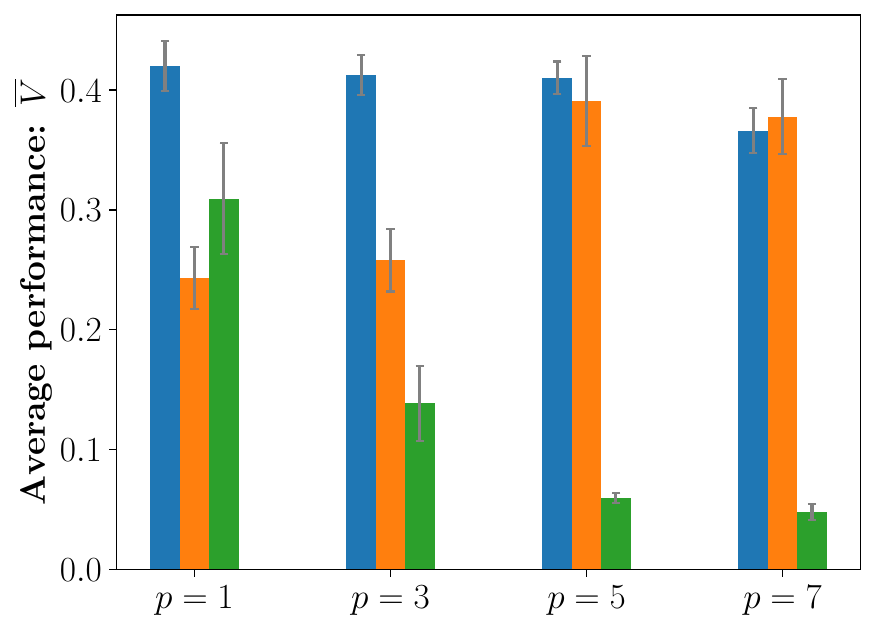}    
        \label{fig:p_change_performance_V}
        \caption[]{\small Average performance}
    \end{subfigure}
    \hspace{0.02\textwidth}
    \begin{subfigure}[t]{0.22\textwidth}   
        \centering 
        \includegraphics[width=\textwidth]{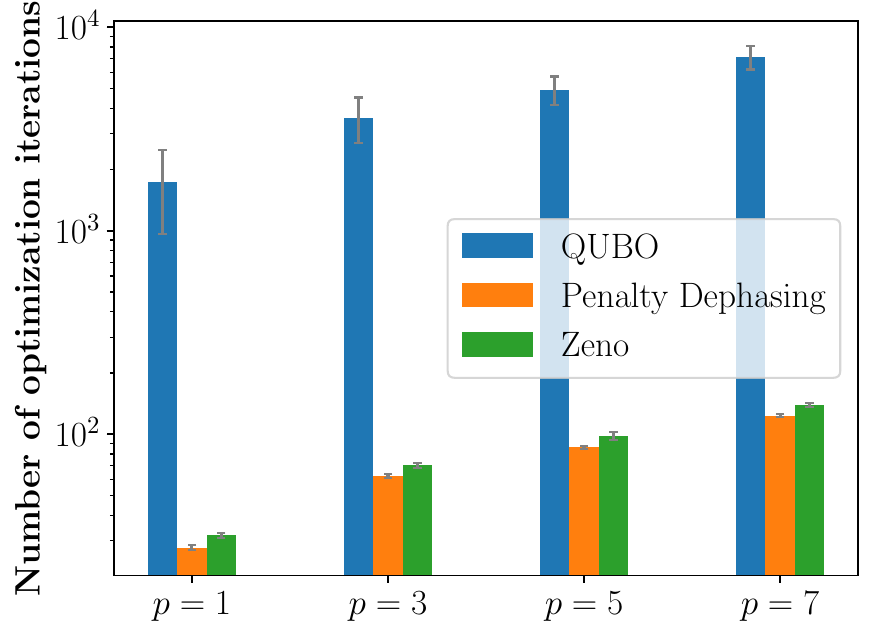}   
        \label{fig:p_change_quantum_resource_nfev}
        \caption[]{\small No. optimization iterations}
    \end{subfigure}
    \caption[]{\small Numerical performance and optimization complexity with varying $p$ and fixed problem size of 5, measure by different metrics: (a) statistical probability of best feasible solution $P_B$; (b) Feasibility ratio $F$; (c) Average performance $V$; and (d) Number of optimization iterations for classical optimizer to converge (y-axis in log scale).} 
    \label{fig:quantum_metrics_performance_varying_p}
\end{figure*}

\section{Results and Analysis}
In this section, we present the results of applying three constraint encoding methods to the Knapsack problem. All experiments were conducted using classical simulation run on commercial laptops.

To evaluate the results, we focus on two key aspects: quantum resource utilization and numerical performance. We analyze the number of ancilla qubits, the number of two-qubit gates per layer, the number of optimization iterations, and the average runtime in seconds for a single problem instance.Three distinct metrics are designed to emphasize different aspects of the results. The first metric, $P_B$, represents the probability of obtaining the best feasible solutions, which is a critical indicator when only the optimal solution is of interest. The second metric, feasibility ratio, is defined as 
$$F = \frac{\sum P(\text{feasible solutions})}{\text{Number of feasible solution}}\times \text{Number of all solutions}$$ The feasibility ratio is particularly useful in scenarios where ensuring feasibility is the primary objective. This metric is designed to account for varying problem sizes and the number of solutions in randomly generated knapsack problems. The final metric is the average performance $$\overline{V} = \frac{\sum_\text{feasible solutions} \text{P(solution)} \times \text{Solution value}}{\text{Best solution value}}$$ This metric provides a more balanced evaluation, capturing good average performance. All evaluations are averaged over 50 randomly generated problem instances, with the shaded error regions representing 25\% of the standard deviation.

\subsection{Sensitivity w.r.t problem size}
We fix the number of layers $p$ at 5. The penalty Lagrange multiplier $P$ for quadratization is set to be $10$ and the penalty weight $\alpha$ for penalty dephasing is set to be $10000$.  Both quantum resources and performance are evaluated as the size of the knapsack problem varies. The problem size is defined by the number of items in the knapsack.

Figure~\ref{fig:quantum_metrics_varying_size} presents the results of quantum resource usage. As expected, the number of ancilla qubits and two-qubit gates increases with the problem size. Quadratization (QUBO) requires fewer ancilla qubits and two-qubit gates compared to the other two methods, as the constraints are embedded into the objective function through penalty terms. In contrast, penalty dephasing and Zeno often involve additional quantum operations to manage constraints during computation, such as dynamically adjusting penalties or performing mid-circuit measurements, respectively.

Although QUBO uses fewer ancilla qubits and two-qubit gates, the results indicate that it requires a greater number of optimization iterations and a longer runtime. A potential explanation for this is that, while QUBO simplifies the quantum circuit by incorporating penalty terms directly into the cost function and reduces the number of qubits, it also complicates the optimization landscape. The optimization process must navigate a more rugged and complex cost function, which may contain numerous local minima due to the penalty terms. Another notable observation is that the average runtime for Zeno is significantly longer than for penalty dephasing, even though Zeno requires fewer optimization iterations. This might be attributed to non-linear mid-circuit measurements which make the states in Zeno circuits mixed states. Simulating mixed states is more time-consuming than simulating pure states, as mixed states are represented by density matrices that occupy a larger computational space, rather than state vectors as in the case of pure states. Consequently, the simulation for Zeno circuits takes longer.

Figure~\ref{fig:performance_varying_size} presents the performance of the three methods. The three metrics illustrate the trade-off between feasible solutions and optimal solutions. Penalty dephasing and Zeno exhibit lower feasibility ratios $F$, indicating that they are less constrained and more likely to find solutions. In contrast, QUBO imposes stronger constraints on the solution space, which likely creates a more rugged and complex optimization landscape, making it more challenging to identify the best solutions. However, the average performance metric suggests that the more constrained QUBO method yields slightly better average performance. This is a reasonable outcome, as most solutions identified by QUBO are feasible, even if they do not have the highest values. These metrics highlight potential differences in the encoding methods, providing insights into their distinct usage scenarios. 

\vspace{-3.2pt}
\subsection{Sensitivity w.r.t number of layers}
We investigate the change of quantum resource usage and performance as the number of layers $p$ changes across the three methods, with the problem size fixed at 5. 
Since the number of ancilla qubits and two-qubit gates are analyzed on a per-layer basis, they will not be included in the following discussion.

Figure~\ref{fig:quantum_metrics_performance_varying_p} illustrates the changes in performance as the number of layers increases. For the probability of finding the best feasible solution ($P_B$), penalty dephasing follows the expectation that as the number of layers increases, the circuit becomes more robust, leading to an increase in $P_B$. However, QUBO and Zeno exhibit the opposite behavior. A possible explanation for this is that as $p$ increases, the optimization landscape becomes more rugged and complex, increasing the likelihood of encountering barren plateaus under the current set of parameters. The suboptimal performance of Zeno may be due to the absence of post-selection. Post-selection was deliberately excluded from the analysis, as it significantly impacts runtime as well as accuracy and we want to control a similar QAOA procedure to compare the encoding methods.

The feasibility ratio for QUBO and penalty dephasing indicates that these methods impose stronger constraints on feasible solutions as $p$ increases. Zeno, on the other hand, continues to show suboptimal results with increasing $p$. The average performance of QUBO remains relatively unchanged, likely because the feasibility ratio and probability of finding the best solution shift in opposite directions as $p$ increases. Penalty dephasing, however, demonstrates the best average performance at 5 layers, which may represent a balance point between finding feasible solutions and identifying the best solutions.

\section{Conclusion}
In conclusion, our comparative analysis of constraint encoding methods for the QAOA provides valuable insights into their effectiveness and practical implications. The sensitivity analysis with respect to problem size and the number of layers revealed key differences in quantum resource utilization and performance metrics.

The QUBO approach demonstrated lower quantum resource requirements, such as fewer ancilla qubits and two-qubit gates, but required more optimization iterations and had longer runtimes. This is likely due to the increased complexity of the optimization landscape from embedding constraints directly into the objective function, making the search for optimal solutions more challenging. Despite these difficulties, QUBO showed slightly better average performance in finding feasible solutions, though it struggled with scalability as problem size increased.

Penalty dephasing improved the probability of finding the best feasible solutions as the number of layers increased but required more quantum resources and didn't always outperform QUBO in feasibility ratio, highlighting a trade-off between resource usage and solution quality. The Quantum Zeno effect, while theoretically promising, encountered practical challenges due to the complexity of frequent mid-circuit measurements, leading to longer runtimes and suboptimal performance, likely from the increased computational burden of simulating mixed states.

Overall, our results suggest that the choice of constraint encoding method should be carefully tailored to the specific characteristics of the problem at hand, considering factors such as problem size, desired accuracy, and available quantum resources. This analysis provides a foundation for future research and practical applications, where these findings can guide the development of more efficient quantum algorithms for complex constraint satisfaction problems. 

\clearpage

\bibliographystyle{ACM-Reference-Format}
\bibliography{acmart}











\end{document}